\documentclass[journal]{IEEEtran}

\usepackage[english]{babel}
\usepackage{cite}
\usepackage{graphicx}
\usepackage[lowtilde]{url}
\usepackage[caption=false,font=footnotesize]{subfig}
\usepackage{amsbsy}
\usepackage{amssymb}
\usepackage{amsmath}
\usepackage{multirow}
\usepackage[euler]{textgreek}
\usepackage{booktabs}
\usepackage{array}
\usepackage[ruled,norelsize]{algorithm2e}
\usepackage{ragged2e}
\usepackage{mathtools}
\usepackage{xcolor}

\makeatletter
\long\def\@makecaption#1#2{\ifx\@captype\@IEEEtablestring%
\footnotesize\begin{center}{\normalfont\footnotesize #1}\\
{\normalfont\footnotesize\scshape #2}\end{center}%
\@IEEEtablecaptionsepspace
\else
\@IEEEfigurecaptionsepspace
\setbox\@tempboxa\hbox{\normalfont\footnotesize {#1.}~~ #2}%
\ifdim \wd\@tempboxa >\hsize%
\setbox\@tempboxa\hbox{\normalfont\footnotesize {#1.}~~ }%
\parbox[t]{\hsize}{\normalfont\footnotesize \noindent\unhbox\@tempboxa#2}%
\else
\hbox to\hsize{\normalfont\footnotesize\hfil\box\@tempboxa\hfil}\fi\fi}
\makeatother

\newcommand{\RN}[1]{%
  \textup{\uppercase\expandafter{\romannumeral#1}}%
}

\makeatletter
\newcommand{\removelatexerror}{\let\@latex@error\@gobble}
\makeatother

\DeclareMathOperator*{\argmax}{arg\,max}

\newcolumntype{C}[1]{>{\centering\let\newline\\\arraybackslash\hspace{0pt}}m{#1}}

\begin{document}
\title{A Knowledge-Driven Quality-of-Experience Model for Adaptive Streaming Videos}
\author{Zhengfang~Duanmu,~\IEEEmembership{Student Member,~IEEE,}
        Wentao~Liu,~\IEEEmembership{Student Member,~IEEE,}\\
        Diqi~Chen,
        Zhuoran~Li,~\IEEEmembership{Student Member,~IEEE,}
        Zhou~Wang,~\IEEEmembership{Fellow,~IEEE,}\\
        Yizhou~Wang,~\IEEEmembership{Member,~IEEE,}
        and~Wen~Gao,~\IEEEmembership{Fellow,~IEEE}
\thanks{Z. Duanmu, W. Liu, Z. Li, and Z. Wang are with the Department of Electrical and Computer Engineering, University of Waterloo, Waterloo, ON N2L 3G1, Canada (e-mail: \{zduanmu, w238liu, z777li, zhou.wang\}@uwaterloo.ca).}
\thanks{D. Chen is with Institute of Computing Technology, Chinese Academy of Sciences, Beijing, 100190, China (e-mail: cdq@pku.edu.cn).}
\thanks{Y. Wang, and W. Gao are with the School of Electronic Engineering and Computer Science, Peking University, Beijing, 100871, China (e-mail: \{yizhou.wang, wgao\}@pku.edu.cn).}
}

\markboth{submitted to IEEE Transactions on Image Processing}%
{Shell \MakeLowercase{\textit{et al.}}: Bare Demo of IEEEtran.cls for Journals}
%




\maketitle

\begin{abstract} The fundamental conflict between the enormous space of adaptive streaming videos and the limited capacity for subjective experiment casts significant challenges to objective Quality-of-Experience (QoE) prediction. Existing objective QoE models exhibit complex functional form, failing to generalize well in diverse streaming environments. In this study, we propose an objective QoE model namely knowledge-driven streaming quality index (KSQI) to integrate prior knowledge on the human visual system and human annotated data in a principled way. By analyzing the subjective characteristics towards streaming videos from a corpus of subjective studies, we show that a family of QoE functions lies in a convex set. Using a variant of projected gradient descent, we optimize the objective QoE model over a database of training videos. The proposed KSQI demonstrates strong generalizability to diverse streaming environments, evident by state-of-the-art performance on four publicly available benchmark datasets.
\end{abstract}


\IEEEpeerreviewmaketitle

\section{Introduction}\label{sec:intro}

\IEEEPARstart{V}{ideo} traffic in various content distribution networks is expected to occupy 71\% of all consumed bandwidth by 2021 and exceed 82\% by 2022~\cite{cisco2017network}. The explosion of data volume introduced by video streaming will quickly drain available network bandwidth in the next decade. Concurrent with the scarcity of network resources is the steady rise in user demands on video quality. With the emergence of new technologies such as 4K, high dynamic range, wide color Gamut, and high frame rate, viewers' expectation on video quality has been higher than ever. The diversity of streaming environments and complexity of human Quality-of-Experience (QoE) response have posed significant challenges to optimal content distribution services.

Adaptive bitrate (ABR) algorithms are the primary tools for modern Internet over-the-top (OTT) video streaming services. In dynamic adaptive streaming environment, ABR achieves player-driven bitrate adaptation by providing video streams in a variety of bitrate and quality levels, and breaking them into small HTTP file segments. Throughout the streaming process, the video player at the client adaptively switches among the available streams by selecting segments based on playback rate, buffer condition and instantaneous throughput, primarily to optimize viewers’ QoE~\cite{jiang2014improving,huang2015buffer,yin2015control,spiteri2016bola,bentaleb2016sdndash,mao2017neural}.

\begin{figure*}[t]
  \centering
  \subfloat[Distribution of VMAF]{\includegraphics[width=0.33\textwidth]{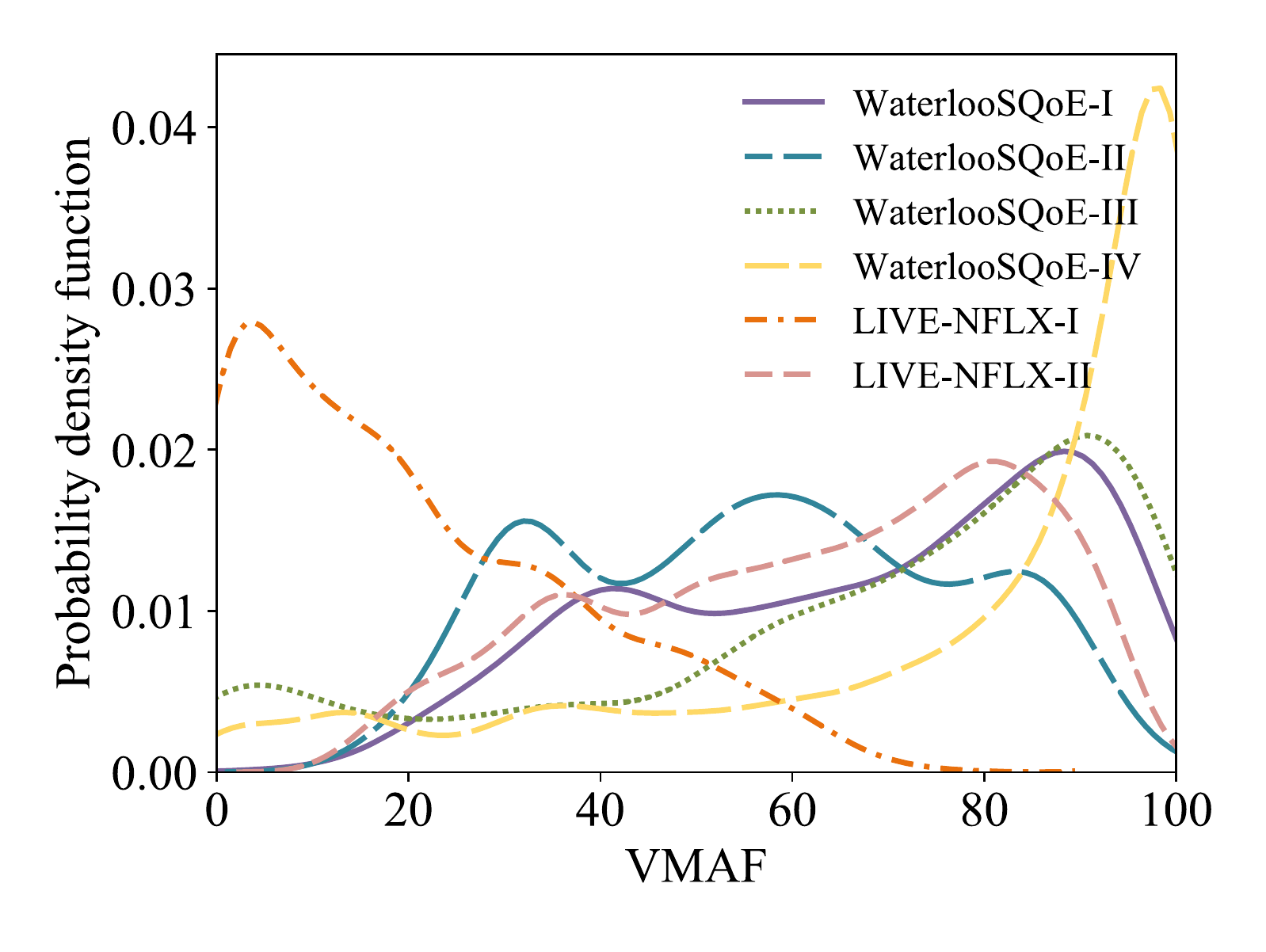}}
  \subfloat[Distribution of rebuffering duration]{\includegraphics[width=0.33\textwidth]{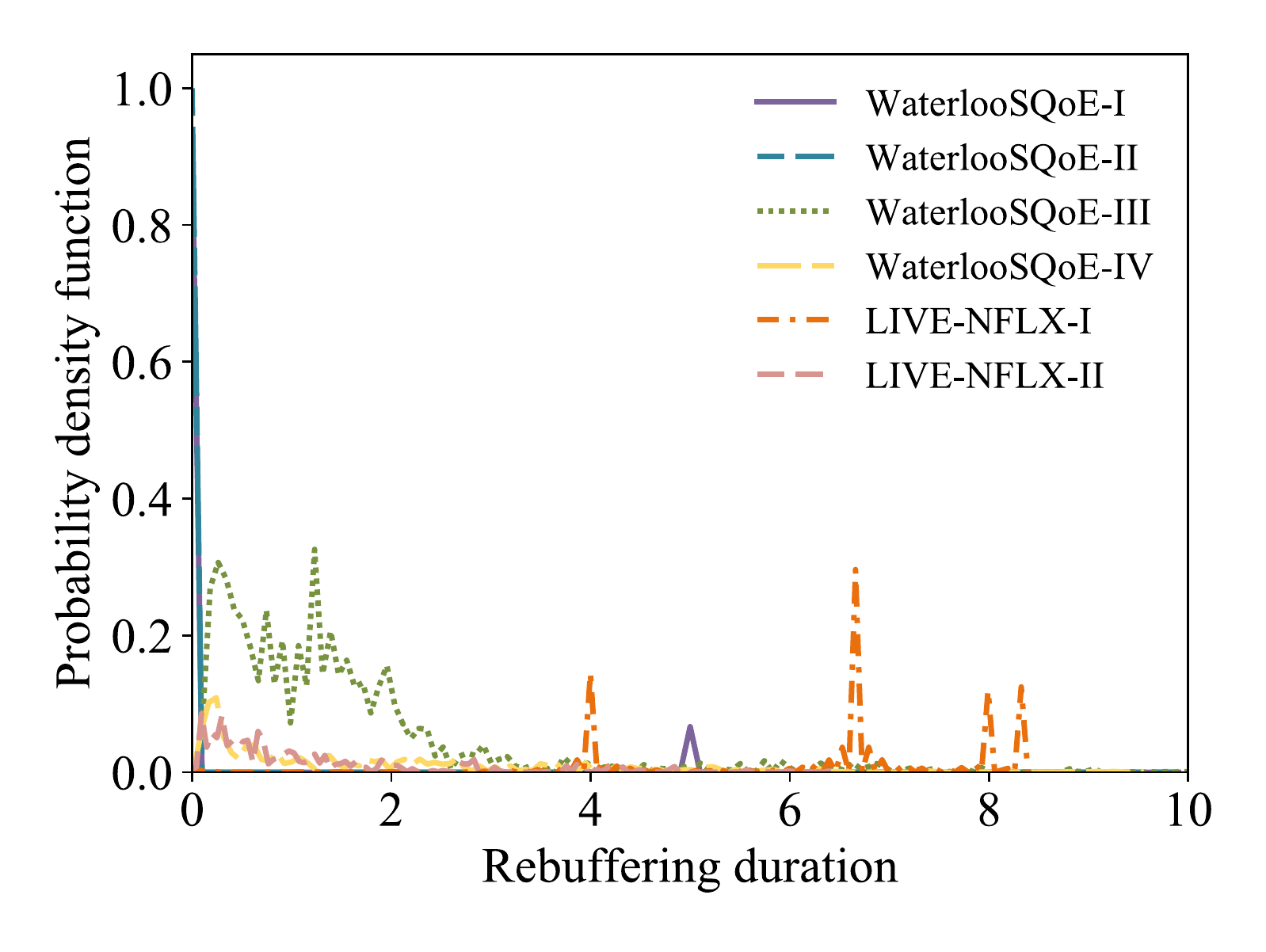}}
  \subfloat[Distribution of adaptation magnitude]{\includegraphics[width=0.33\textwidth]{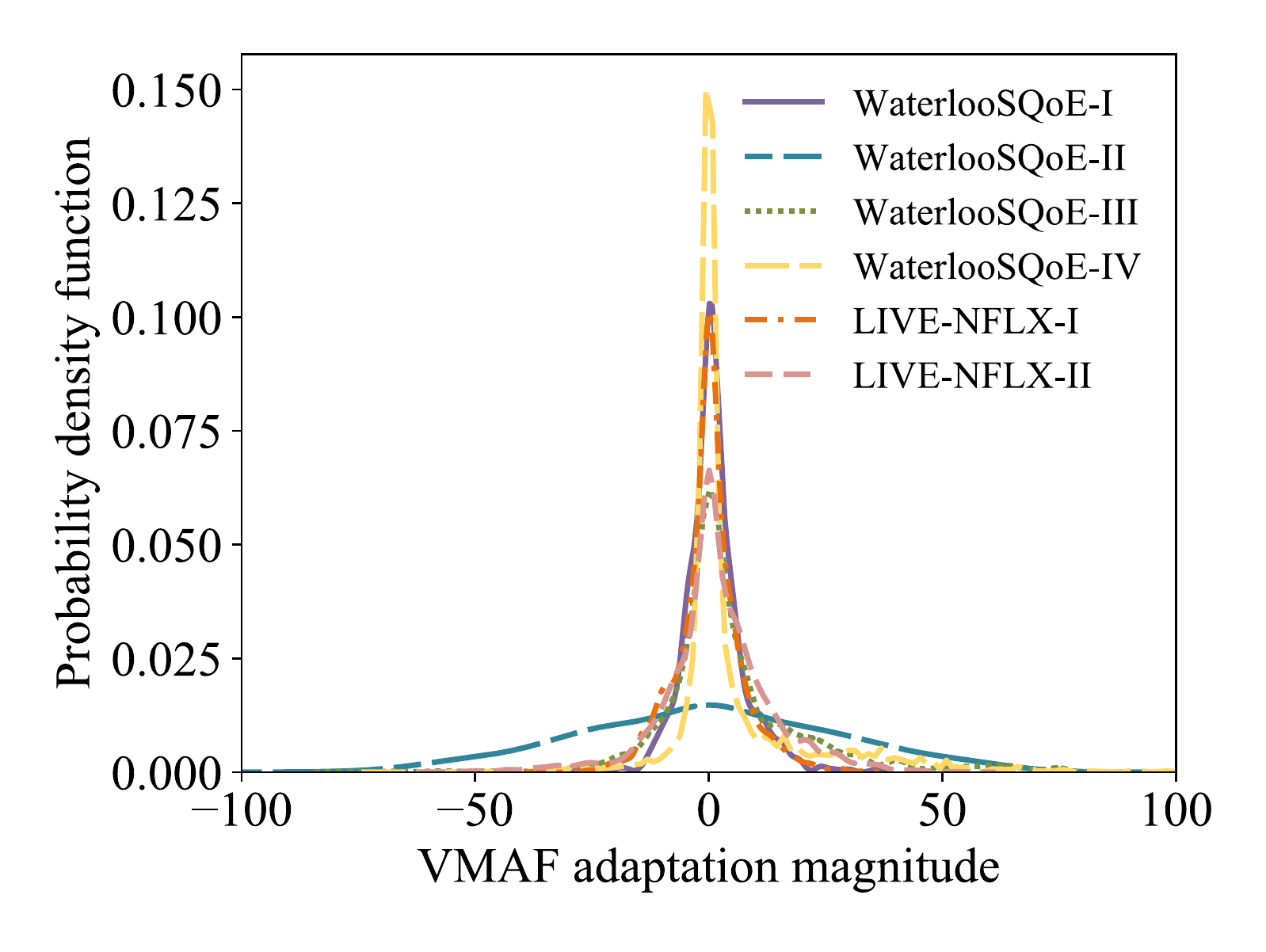}}
  \caption{There exists significant variance on the characteristics of streaming videos, evident by the distributions of (a) VMAF, (b) rebuffering duration, and (c) adaptation magnitude in six publicly available datasets.}\label{fig:feature_pdf}
\end{figure*}

With many ABR algorithms at hand, it becomes pivotal to measure their performance so as to guide the network resource allocation. Therefore, the development of an accurate objective QoE model lies in the root of ABR systems. State-of-the-art QoE models employ sophisticated machine learning techniques such as random forest~\cite{breiman2001random}, support vector machine~\cite{burges1998tutorial}, and neural network~\cite{hagan1996neural} to combat the complex human visual system (HVS). The success of the approach heavily depends on the quantity and quality of training data, both of which are extremely limited in practice. First, there is a major conflict between the enormous sample space and the limited capacity for subjective QoE measurements. For example, the number of possible adaptation patterns in a $d$-temporal segment and $b$-encoding level streaming video is $b^d$, which further expands with respect to the source content, encoder type, and rebuffering patterns. By contrast, collecting group-truth data via subjective testing is expensive and time-consuming. The largest publicly available subject-rated QoE database contains only $1,350$ samples~\cite{duanmu2019sqoe4}, which are deemed to be extremely sparsely distributed in the sample space. Second, the learning-based models assume that the training samples and testing samples come from the same distribution. However, the assumption has never been justified in the existing studies and may hardly hold in practice. A motivating example is shown in Fig.~\ref{fig:feature_pdf}, where the probability density functions of video presentation quality measured by a state-of-the-art VQA model VMAF~\cite{li2016VMAF}, rebuffering duration, and quality adaptation magnitude in six publicly available streaming QoE datasets are presented. Clearly, there is significant variability on the characteristics of streaming videos across different datasets, suggesting that an objective QoE model optimized on a simple dataset such as WaterlooSQoE-I~\cite{duanmu2016sqi} may yield very poor predictions on complex datasets as WaterlooSQoE-III~\cite{duanmu2018quality}, WaterlooSQoE-IV~\cite{duanmu2019sqoe4}, and LIVE-NFLX-II~\cite{bampis2018towards}, and vice versa. The streaming video probability density estimation is further complicated by the concept drift problem~\cite{gama2014survey}, where the characteristics of streaming video changes over time. For example, the drift in streaming video distribution may arise from the advancement of video acquisition~\cite{kang2003high,ng2005light,ishii20102000}, compression~\cite{netflix2015bitrateladder,toni2015optimal,de2016complexity}, transmission~\cite{jiang2014improving,huang2015buffer,yin2015control,spiteri2016bola,bentaleb2016sdndash,mao2017neural}, and reproduction systems~\cite{nasiri2015perceptual,wang2017begin,li2019avc}, and the steady rise in viewers' expectation on video quality~\cite{oliver1980cognitive,duanmu2018ect}. Consequently, the construction of a large-scale and representative training dataset remains an elusive goal.

Hindered by the two fundamental conflicts, it is highly difficult to develop objective QoE models that generalize to diverse impairments. To this end, we propose an objective QoE model namely knowledge-driven streaming quality index (KSQI) to integrate prior knowledge of the HVS and a limited number of training data. From a Bayesian perspective, we show one possible solution to the QoE assessment problem resides in a deeper understanding of the HVS.

Given a collection of subjective QoE studies, how do we make use of the knowledge in a principled and scalable manner? To answer the question, we analyze the HVS properties observed from existing subjective QoE studies, from which we derive a system of linear inequalities. We further show that a family of objective QoE models lies within a convex set that results from the intersection of a hyper-plane and a positive cone in a functional space. This gives us both guidance on the form of our model as well as constraints.

Building upon insights of HVS properties, how do we design an objective QoE model that can accurately predict subjective QoE response? We demonstrate that the QoE model parameter estimation problem can be formulated as a quadratic programming problem. Using a variant of projected gradient descent, we optimize the proposed model over a database of training samples with limited adaptation patterns. The resulting model is computationally efficient, mathematically well-behaved and perceptually grounded.

We compare KSQI to ten objective QoE models on four benchmark datasets covering a broad set of video contents, encoder configurations, network conditions, ABR algorithms, and viewing devices. KSQI shows very strong generalizability in all considered scenarios, significantly outperforming all existing schemes. We show that the proposed model is not only superior on average, but also in extreme cases via a set of intuitive examples. We have made the implementation of all objective QoE models available at \url{https://github.com/zduanmu/ksqi} to facilitate future objective QoE research.

In summary, this paper makes the following key contributions:
\begin{itemize}
    \item Mathematical analysis on the space of QoE functions for adaptive streaming videos;
    \item Design of an objective QoE model combining the constraints from our analysis and human annotated data;
    \item By far the most comprehensive evaluation of objective QoE models.
\end{itemize}

\section{Related Works}\label{sec:pwork}
Over the past decade, QoE for adaptive streaming videos has been a rapidly evolving research topic and has attracted an increasing amount of attention from both industry and academia. Existing objective QoE models can be categorized based on the observation space and the functional form that they rely on to make QoE predictions. The earliest QoE models simply correlate the statistics of rebuffering events to QoE~\cite{watanabe2007objective}, assuming rebuffering dominates the viewing experience. Although several efforts have put forth to improve the prediction accuracy~\cite{mok2012qdash,tobias2013youtube}, the negligence of picture quality reduces the relevance of the QoE models to real-world ABR scenarios. To overcome the limitations of the rebuffering-centric QoE models, many studies propose to complement rebuffering duration with average bitrate or quantization parameter (QP) as the input to the quality prediction model. These models compute the QoE as a weighted average of bitrate/QP and rebuffering duration, where the trade-off parameter is either determined empirically~\cite{liu2012case} or by solving a regression problem~\cite{xue2014assessing}. Motivated by observations that frequent quality adaptations annoy viewers~\cite{ni2011flicker,rehman2013perceptual,duanmu2017qoe,duanmu2018ect}, a number of models has been developed to explicitly quantify the quality adaptation experience and then linearly combine it with average bitrate and rebuffering duration~\cite{rehman2013perceptual,yin2015control,spiteri2016bola}. Due to the simplicity, these models remain the standard criterion for the assessment and optimization of ABR systems~\cite{yin2015control,spiteri2016bola,mao2017neural,akhtar2018oboe}. Despite the demonstrated success, the aforementioned QoE models assume every single bit contributes equally to the video quality. However, the assumption is fundamentally flawed according to the rate-distortion theory~\cite{berger1971rate}, and may deteriorate in different compression, transmission and reproduction systems~\cite{de2016complexity,wang2016method,wang2017begin}. Recent efforts suggest to replace bitrate by state-of-the-art video quality assessment (VQA) models~\cite{pinson2010vqm,rehman2015ssimplus,li2016VMAF} as the presentation quality measure~\cite{duanmu2016sqi,liu2015deriving,bentaleb2016sdndash}, achieving highly competitive performance on existing benchmarks. All these models make a priori assumptions about the form of the QoE models. The subjective QoE response with respect to rebuffering and quality adaptation, however, can vary significantly from an exponential or logarithmic function.

Supposing subjective QoE response is too complex to model with simple parametric functions, recent objective QoE models utilize machine learning techniques such as non-linear auto-regressive model~\cite{bampis2017continuous}, neural network~\cite{singh2012quality}, support vector machine~\cite{bampis2017atlas}, and random forest~\cite{itu2017pnats,duanmu2018ect} to map streaming video features to subjective opinion scores. Although these models can fit arbitrary complex continuous functions~\cite{hornik1989multilayer}, they are often susceptible to overfitting. The defect is exaggerated by the limited capacity for subjective experiment and the concept drift problem as will be demonstrated in the subsequent sections.

In addition to the specific limitations the two kinds of models may respectively have, they may qualitatively contradict the HVS properties observed from the existing subjective studies. Furthermore, the objective QoE models have not received comprehensive evaluation on subject-rated datasets of diverse video contents, encoder types, network conditions, ABR algorithms, and viewing devices. While many recent works acknowledge the importance of objective QoE models~\cite{huang2015buffer,yin2015control,mao2017neural,akhtar2018oboe,huang2018qarc}, a careful analysis, modeling, and evaluation of the models has yet to be done. We wish to address this void. In doing so, we seek a good compromise between 1) simple and rigid model building upon intuitive understanding of QoE and 2) complex and indefinite models requiring a significant number of training samples.

\section{Characterizing the QoE Functions}\label{sec:qoe_space}
Despite the complexity of HVS, there are three widely accepted key influencing factors in QoE of streaming videos: video presentation quality, rebuffering, and quality adaptation (switching between profiles)~\cite{seufert2014survey,garcia2014quality,yin2015control,spiteri2016bola,bampis2017qoe,duanmu2018quality,bampis2018towards}. The simplest approach to parametrizing the QoE function is to assume the three elements are additive. Formally, the QoE of chunk $t$ is determined by
\begin{equation*}
    Q_t = P_t + S_t + A_t,
\end{equation*}
where $P_t$, $S_t$, and $A_t$ denote the presentation quality, the rebuffering QoE function, and the adaptation QoE function of chunk $t$, respectively. For simplicity, we will drop the subscript $t$ from $P_t$ and $Q_t$, denote $P_{t-1}$ by lower case $p$, and $P_{t} - P_{t-1}$ by $\Delta p$ in the rest of this section unless otherwise specified.

While the presentation quality $P$ can be obtained from subjective tests or a reliable VQA model, the two QoE functions $S$ and $A$ are not well studied. Defining the space of QoE functions helps us build a model of these functions. It not only guides us as to the form such a model should take, but also determines the constraints these functions must satisfy. We begin by summarizing observations from a collection of existing subjective QoE studies, and then formulate the domain knowledge to define the space of these functions.

\subsection{Space of Rebuffering Experience Function $S$}
First, we assume that the influence of each rebuffering event is independent, additive, and only determined by the previous chunk's presentation quality $p$ and the rebuffering duration $\tau$~\cite{duanmu2016sqi, bampis2017qoe, ma2019gmad}. This assumption allows us to analyze each rebuffering event separately, and reduces dimensions of the functional space~\cite{duanmu2016sqi}. As such, $S$ can be modeled as a bi-variate function $S(p,\tau) \in \{S|S:\mathbb{R}^2\rightarrow \mathbb{R}\}$.

Second, various subjective tests~\cite{tobias2011youtube, dobrian2011understanding} have attested that rebuffering duration is negatively correlated with the overall QoE of streaming videos, while very short rebuffering may not be perceived and thus has little impact on QoE~\cite{qi2006effect,staelens2010assessing}. Formally,
\begin{equation}\label{eq:s1}
    \left\{
    \begin{array}{ll}
        S(p, \tau_1) \geq S(p, \tau_2),  &\forall p, \tau_1 \leq \tau_2 \\
        S(p, 0) = 0,  &\forall p
    \end{array}\right..
\end{equation}

The third assumption is that the QoE drop tends to be greater when the presentation quality of the previous chunk is higher, \textit{i.e.}
\begin{equation}\label{eq:s2}
S(p_1, \tau) \geq S(p_2, \tau), \forall p_1 \leq p_2, \tau.
\end{equation}
Such a trend has been observed in recent subjective tests~\cite{duanmu2016sqi, bampis2017qoe}, and may be explained by the expectation confirmation theory~\cite{oliver1980cognitive}.

The fourth assumption is elicited from the fact that, given a constant presentation quality and a fixed total duration of rebuffering, the overall QoE degrades as the number of rebuffering occurrences increases~\cite{moorthy2012video, tobias2013youtube, pastrana2004sporadic}. Mathematically, this may be expressed as
\begin{equation}\label{eq:s3}
S(p, \tau_1) + S(p, \tau_2) \leq S(p, \tau_1 + \tau_2), \forall p, \tau_1, \tau_2.
\end{equation}

The fifth remark on $S$ is that, given the same rebuffering duration, videos with higher presentation quality consistently deliver higher overall QoE, despite the greater penalty for the rebuffering event~\cite{ma2019gmad}. This statement can be formulated as
\begin{equation}\label{eq:s4}
S(p_1, \tau) + p_1 \leq S(p_2, \tau) + p_2, \forall p_1 \leq p_2, \tau.
\end{equation}

In summary, we define the theoretical space of rebuffering QoE functions $S$ as
\begin{equation}\label{eq:space_S}
\resizebox{0.91\hsize}{!}{
    $\begin{array}{ll}
    \mathcal{W}_S \coloneqq & \{S:\mathbb{R}^2 \rightarrow \mathbb{R}| S(p,0) = 0, S(p, \tau_1) \geq S(p, \tau_2),\\
    & S(p_1, \tau) \geq S(p_2, \tau), S(p, \tau_1) + S(p, \tau_2) \leq S(p, \tau_1 + \tau_2), \\
    & S(p_1, \tau) + p_1 \leq S(p_2, \tau) + p_2, \forall p, \tau, \tau_1 \leq \tau_2, p_1 \leq p_2\}. 
    \end{array}$
    }
\end{equation}
The equality constraint and inequality constraints in~\eqref{eq:space_S} represent a hyperplane and a positive cone, respectively~\cite{bishop2006pattern}. The convexity of hyperplane and cone determines that their intersection is also convex.

\subsection{Space of Adaptation Experience Function $A$}
Similar to video rebuffering, we first assume that the influence of each quality adaptation event is independent and additive, and only depends on the instantaneous presentation quality $p$, and the intensity of the adaptation $\Delta p$~\cite{ni2011flicker, rehman2013perceptual, moorthy2012video, duanmu2017qoe}. Thus the adaptation QoE function $A$ lies in the space defined by $\{A=A(p, \Delta p)|A: \mathbb{R}^2 \rightarrow \mathbb{R}\}$. 

The second assumption is an intuitive one that $A$ must have the same sign as the quality adaptation $\Delta p$~\cite{grafl2013representation, moorthy2012video, rehman2013perceptual, duanmu2017qoe}. This assumption suggests that people always assign a penalty to presentation quality degradation, reward to quality elevation, and neither penalty nor reward when no quality adaptation occurs. Mathematically, the assumption can be expressed as
\begin{equation}\label{eq:a1}
    \left\{
    \begin{array}{ll}
         A(p, \Delta p) < 0, & \forall p, \Delta p < 0 \\
         A(p, \Delta p) > 0, & \forall p, \Delta p > 0 \\
         A(p, 0) = 0, & \forall p
    \end{array}\right..
\end{equation}

Further analysis~\cite{ni2011flicker, moorthy2012video, rehman2013perceptual, duanmu2017qoe} on the relationship between the QoE adjustment $A$ and the intensity of quality adaptation $\Delta p$ indicates that subjects tend to give greater QoE penalty or reward when quality drops or improves by a greater amount. This finding, together with the second assumption, prompts our third assumption: $A$ is monotonically increasing with regards to $\Delta p$. Formally, the two assumptions combined can be represented by
\begin{equation}\label{eq:a2}
    A(p, \Delta p_1) \leq A(p, \Delta p_2), \forall p, \Delta p_1 \leq \Delta p_2.
\end{equation}

Experiments in~\cite{duanmu2017qoe} find that quality degradation occurring in the high quality range leads to greater amount of penalty than that occurring in the low quality range, while quality elevation in the high quality range results in smaller rewards. Such an observation leads to the fourth assumption that the function $A$ is monotonically decreasing along the other axis $p$, \textit{i.e.}
\begin{equation}\label{eq:a3}
A(p_1, \Delta p) \geq A(p_2, \Delta p), \forall p_1 \leq p_2, \Delta p.
\end{equation}

Another commonly observed trend in adaptation QoE is that the reward for a positive quality change is relatively smaller than the penalty for a negative one given the same intensity of quality adaptation and the same average presentation quality~\cite{ni2011flicker, rehman2013perceptual, duanmu2017qoe}. Formally, this can be summarized by
\begin{equation}\label{eq:a4}
A(p, -\Delta p) + A(p - \Delta p, \Delta p) \leq 0, \forall p, \Delta p.
\end{equation}

In summary, we define the space of adaptation experience functions $A$ as
\begin{equation}\label{eq:space_A}
\resizebox{0.89\hsize}{!}{
    $\begin{array}{ll}
     \mathcal{W}_A \coloneqq & \{A: \mathbb{R}^2 \rightarrow \mathbb{R}| A(p, 0) = 0, A(p, \Delta p_1) \leq A(p, \Delta p_2),\\
     & A(p, -\Delta p) + A(p-\Delta p, \Delta p) \leq 0, \\
    & A(p_1, \Delta p) \geq A(p_2, \Delta p),  \forall p, \Delta p, p_1 \leq p_2, \Delta p_1 \leq \Delta p_2\}.
    \end{array}$
    }
\end{equation}
Analogous to the rebuffering experience function, the adaptation experience function $A$ also lies in a convex set.

\section{A Knowledge-Driven QoE Model}\label{sec:oqoe}
Given the functional space constrained by the hyperplane and the positive cone described above, there are infinite number of functions lying in the space. A good QoE model should be in close agreement with human perception. In this section, we present the roadmap to design a perceptually grounded objective QoE model.

\subsection{Modeling the Presentation Quality}
Traditionally, for the sake of operational convenience, bitrate is often used as the major indicator of video presentation quality~\cite{huang2015buffer,mok2012qdash,li2014probe,yin2015control,spiteri2016bola,mao2017neural}. However, bitrate may heavily deviate from perceptual quality. The presentation quality model should provide meaningful and consistent QoE predictions across video contents, video resolutions, and viewing conditions/devices. To the best of our knowledge, currently the only video QoE models that satisfy such requirements are SSIMplus~\cite{rehman2015ssimplus} and VMAF~\cite{li2016VMAF}. Both models perform consistently well on various subject-rated video databases~\cite{liu2018bvqa,bampis2018simple}, making them an appropriate component in KSQI. In the rest of the paper, we present our results using VMAF as our presentation quality model as it is open source that facilitates reproducible research. Although the presentation quality scores are not available to the adaptive streaming player by default, they can either be embedded into the manifest file that describes the specifications of the video, or carried in the metadata of the video container. Thanks to the light overhead, the feature embedding technique has been successfully deployed in practical QoE measurement~\cite{duanmu2016sqi,wang2017method} and ABR optimization systems~\cite{bentaleb2016sdndash,wang2016method}.


\subsection{Modeling the Rebuffering QoE Function $S$}\label{sec:modeling_s}
Strictly speaking, $\mathcal{W}_S$ is a space of continuous functions, but we may approximate it in terms of a vector space by densely sampling the supporting domain of $S$. Specifically, the supporting domain of $S$ is defined as $\{(p, \tau)|p \in [0, P], \tau \in [0, \tau_{\max}]\}$, where $P$ indicates the best quality, and $\tau_{\max}$ is the maximum rebuffering duration. By uniformly sampling both $p$ and $\tau$, we approximate the function $S$ with a finite-size matrix $\mathbf{S}\in \mathbb{R}^{(N+1) \times (N+1)}$, where an element $s_{i,j}$ denotes the QoE penalty when $(p, \tau) = \left(\frac{i-1}{N}P, \frac{j-1}{N}\tau_{\rm{max}} \right)$. We then vectorize $\mathbf{S}$ as $\mathbf{s} \in \mathbb{R}^{(N+1)^2}$ for the convenience of further formulation. Finally, we are able to approximate the functional space $\mathcal{W}_S$ with a vector space
\begin{equation*}
    \mathcal{W}_\mathbf{s} \coloneqq \{\mathbf{s} \in \mathbb{R}^{(N+1)^2} | \mathbf{G}^s\mathbf{s} \leq \mathbf{h}^s, \mathbf{B}^s\mathbf{s} = \mathbf{c}^s\}, \label{eq:appr_space_S}
\end{equation*}
where $\mathbf{G}^s, \mathbf{h}^s, \mathbf{B}^s$ and $\mathbf{c}^s$ are constructed so that all the entries in $\mathbf{s}$ should satisfy the constraints in \eqref{eq:space_S}.

\begin{table*}[t]
\centering
\caption{Comparison of Objective QoE Models}
\label{tab:existingQoE}
  \begin{tabular}{c|c c|c c|c}
  \toprule
      \multirow{2}{*}{QoE model} & \multicolumn{2}{c|}{Presentation quality}      & \multicolumn{2}{c|}{Rebuffering} & Switching  \\\cline{2-6}
                                  & Regression function & Features & Regression function  & Features & Regression function  \\\hline
      Mok2011~\cite{mok2012qdash}               & ---    & ---     & linear & $\tau$ & --- \\
      FTW~\cite{tobias2013youtube}              & ---    & ---     & exponential & $\tau$ & --- \\
      Liu2012~\cite{liu2012case}                & linear & bitrate & linear & $\tau$ & --- \\
      Xue2014~\cite{xue2014assessing}           & linear & QP      & logarithmic & $\tau$ & --- \\
      Yin2015~\cite{yin2015control}             & linear & bitrate & linear & $\tau$ & linear \\
      Spiteri2016~\cite{spiteri2016bola}        & logarithmic & bitrate & linear & $\tau$ & logarithmic \\
      Bentaleb2016~\cite{bentaleb2016sdndash}   & linear & VQA     & linear & $\tau$ & --- \\
      SQI~\cite{duanmu2016sqi}                  & ---    & VQA     & linear & VQA, $\tau$ & --- \\
      P.1203~\cite{itu2017pnats}                & random forest & bitrate, resolution & random forest & $\tau$ & random forest \\
      VideoATLAS~\cite{bampis2017atlas}         & SVR    & VQA     & SVR    & $\tau$ & SVR \\ \hline
      KSQI                                      & ---    & VQA     & non-parametric & VQA, $\tau$ & non-parametric \\
  \bottomrule
  \end{tabular}
\end{table*}

Given a training set of $M_s$ video sequences, each of which has $C_s$ chunks, one or more rebuffering events, no adaptation, and a mean opinion score (MOS) $Q_m$ to indicate its overall QoE, we want to obtain a vector $\mathbf{s}^*\in \mathcal{W}_\mathbf{s}$ that best fits the training data. Besides, it is beneficial to impose smoothness on the function $S$. Practically, many subjective experiments have empirically shown the smoothness of the QoE functions~\cite{duanmu2016sqi, bampis2017continuous}. Mathematically, smoothness regularization may lead to well-behaved solutions. Thus, the objective function of $\mathbf{s}$ can be defined as
\begin{equation}
    \begin{aligned}
    L_\mathbf{s} \coloneqq \epsilon^{\textrm{F}}_\mathbf{s} + \lambda \epsilon^{\textrm{S}}_\mathbf{s},
    \end{aligned}
    \label{eq:obj_s}
\end{equation}
where $\lambda > 0$ is a weighting factor. We adopt the mean squared error as the fidelity term $\epsilon^{\textrm{F}}_\mathbf{s}$, and the sum of squared second-order differences along $i$ and $j$ axes as the smoothness term $\epsilon^{\textrm{S}}_\mathbf{s}$. Formally, we define that
\begin{equation}
    \left\{
    \begin{array}{l}
    \epsilon^{\textrm{F}}_\mathbf{s} \coloneqq\frac{1}{M_s}\sum_{m=1}^{M_s}\left[Q_m - \frac{1}{C_s}\sum_{c=1}^{C_s}(P_{m_c} + s_{i_{m_c}, j_{m_c}}) \right]^2 \\
    \epsilon^{\textrm{S}}_\mathbf{s}  \coloneqq \frac{1}{(N+1)^2}\sum_{i=1}^{N+1}\sum_{j=1}^{N+1}\left[\left(\frac{\partial^2 s_{i,j}}{\partial i^2}\right)^2 + \left(\frac{\partial^2 s_{i,j}}{\partial j^2}\right)^2\right]
    \end{array}\right., \nonumber
\end{equation}
where $P_{m_c}$ and $s_{i_{m_c}, j_{m_c}}$ denote the presentation quality and rebuffering QoE penalty at chunk $c$ of video $m$, respectively. It is not hard to see that both $\epsilon^{\textrm{F}}_\mathbf{s}$ and $\epsilon^{\textrm{S}}_\mathbf{s}$ take quadratic forms of $\mathbf{s}$. As a result, we are able to estimate the rebuffering QoE matrix $\mathbf{S}$ by solving the following quadratic programming problem
\begin{equation}\label{eq:qp_s}
    \begin{array}{ll}
    \underset{\mathbf{s}}{\text{minimize}}
& L_\mathbf{s} = \epsilon^{\textrm{F}}_\mathbf{s} + \lambda \epsilon^{\textrm{S}}_\mathbf{s} \\
\text{subject to}
& \mathbf{s} \in \mathcal{W}_\mathbf{s}.
    \end{array}
\end{equation}
The convexity of $\mathcal{W}_\mathbf{s}$ and~\eqref{eq:obj_s} implies that there exists a unique solution for the optimization problem. The problem can be efficiently solved with projected gradient descent-based algorithms such as alternating direction method of multipliers~\cite{boyd2011distributed}. 

\subsection{Modeling the Adaptation QoE Function $A$}
Following the same approach, we work with the discrete version of of $A$. The supporting domain of $A$ is $\{(p, \Delta p)|p \in [0, P], \Delta p \in [-p, P - p]\}$, since the presentation quality could go neither below $0$ nor over the best quality $P$. By uniformly sampling both $p$ and $\Delta p$, we approximate the function $A$ with a finite-size matrix $\mathbf{A}\in \mathbb{R}^{(N+1) \times (N+1)}$, where an entry $a_{i,j}$ denotes the QoE change when $(p, \Delta p) = \left(\frac{i-1}{N}P, \frac{j-i}{N}P \right)$, and then vectorize $\mathbf{A}$ as $\mathbf{a} \in \mathbb{R}^{(N+1)^2}$. Finally, the vector space of adaptation experience function becomes
\begin{equation*}
    \mathcal{W}_\mathbf{a} \coloneqq \{\mathbf{a} \in \mathbb{R}^{(N+1)^2} | \mathbf{G}^a\mathbf{a} \leq \mathbf{h}^a, \mathbf{B}^a\mathbf{a} = \mathbf{c}^a\}, \label{eq:appr_space_A}
\end{equation*}
where $\mathbf{G}^a, \mathbf{h}^a, \mathbf{B}^a$ and $\mathbf{c}^a$ are according to the constraints in \eqref{eq:space_A}.

Given a training set of $M_a$ video sequences, each of which has $C_a$ chunks, no rebuffering events, and a MOS $Q_m$, we aim to optimize
\begin{equation*}
    \begin{aligned}
    L_\mathbf{a} \coloneqq \epsilon^{\textrm{F}}_\mathbf{a} + \lambda \epsilon^{\textrm{S}}_\mathbf{a},
    \end{aligned}
    \label{eq:obj_a}
\end{equation*}
where
\begin{equation}
    \left\{
    \begin{array}{l}
    \epsilon^{\textrm{F}}_\mathbf{a} \coloneqq\frac{1}{M_a}\sum_{m=1}^{M_a}\left[Q_m - \frac{1}{C_a}\sum_{c=1}^{C_a}(P_{m_c} + a_{i_{m_c}, j_{m_c}}) \right]^2 \\
    \epsilon^{\textrm{S}}_\mathbf{a}  \coloneqq \frac{1}{(N+1)^2}\sum_{i=1}^{N+1}\sum_{j=1}^{N+1}\left[\left(\frac{\partial^2 a_{i,j}}{\partial i^2}\right)^2 + \left(\frac{\partial^2 a_{i,j}}{\partial j^2}\right)^2\right]
    \end{array}\right.. \nonumber
\end{equation}
Here, $s_{i_{m_c}, j_{m_c}}$ denotes the quality adaptation experience at chunk $c$ of video $m$. The optimal quality adaptation experience matrix $A$ can be obtained by solving the following quadratic programming problem
\begin{equation}\label{eq:qp_a}
    \begin{array}{ll}
    \underset{\mathbf{a}}{\text{minimize}}
& L_\mathbf{a} = \epsilon^{\textrm{F}}_\mathbf{a} + \lambda \epsilon^{\textrm{S}}_\mathbf{a} \\
\text{subject to}
& \mathbf{a} \in \mathcal{W}_\mathbf{a}.
    \end{array}
\end{equation}

\subsection{Overall QoE}
In practice, one usually requires a single end-of-process QoE measure. We use the mean value of the predicted QoE over the whole playback duration to evaluate the overall QoE. To reduce the memory usage, the end-of-process QoE can be computed in a moving average fashion
\begin{equation}
\nonumber Y_t = \frac{(t - 1)Y_{t-1} + Q_t}{t},
\end{equation}
where $Y_t$ is the cumulative QoE up to the $t$-th segment in the streaming session.

\section{Experiments}
In this section, we first describe the experimental setups and evaluation criteria. We then compare KSQI with classic and state-of-the-art objective QoE models. Furthermore, we develope an efficient methodology for examining the best-case performance of objective QoE models. Finally, we conduct a series of ablation experiments to identify the contributions of the core factors in KSQI.
\begin{table*}[t]
\centering
    \caption{PLCC between the objective QoE model prediction and MOS on the benchmark datasets.}\label{tab:plcc}
    \begin{tabular}{c|c c c c|c c}
    \toprule
      QoE model & LIVE-NFLX-I & LIVE-NFLX-II & WaterlooSQoE-III & WaterlooSQoE-IV & Average & Weighted Average\\ \hline
      Mok2011~\cite{mok2012qdash}               & 0.292 & 0.512 & 0.173 & 0.046 & 0.256 & 0.166 \\
      FTW~\cite{tobias2013youtube}              & 0.286 & 0.568 & 0.323 & 0.147 & 0.331 & 0.263 \\
      Xue2014~\cite{xue2014assessing}           & ---   & 0.788 & 0.387 & 0.166 & 0.447 & 0.328 \\
      Liu2012~\cite{liu2012case}                & 0.524 & 0.732 & 0.609 & 0.282 & 0.537 & 0.438 \\
      Yin2015~\cite{yin2015control}             & 0.376 & 0.673 & 0.722 & 0.323 & 0.524 & 0.466 \\
      VideoATLAS~\cite{bampis2017atlas}         & 0.100 & 0.644 & 0.385 & 0.675 & 0.451 & 0.586 \\
      P.1203~\cite{itu2017pnats}                & 0.325 & 0.817 & 0.769 & 0.636 & 0.637 & 0.679 \\
      Bentaleb2016~\cite{bentaleb2016sdndash}   & 0.741 & 0.898 & 0.625 & 0.682 & 0.737 & 0.713 \\
      Spiteri2016~\cite{spiteri2016bola}        & 0.612 & 0.731 & \textbf{0.809} & 0.685 & 0.709 & 0.714 \\
      SQI~\cite{duanmu2016sqi}                  & \textbf{0.756} & \textbf{0.910} & 0.673 & \textbf{0.717} & \textbf{0.764} & \textbf{0.745} \\ \hline
      KSQI                                      & \textbf{0.753} & \textbf{0.905} & \textbf{0.794} & \textbf{0.720}  & \textbf{0.793} & \textbf{0.769} \\
    \bottomrule
    \end{tabular}
\end{table*}

\begin{table*}[t]
\centering
    \caption{SRCC between the objective QoE model prediction and MOS on the benchmark datasets.}\label{tab:srcc}
    \begin{tabular}{c|c c c c|c c}
    \toprule
      QoE model & LIVE-NFLX-I & LIVE-NFLX-II & WaterlooSQoE-III & WaterlooSQoE-IV & Average & Weighted Average\\ \hline
      Mok2011~\cite{mok2012qdash}               & 0.335 & 0.516 & 0.152 & 0.056 & 0.265 & 0.171 \\
      FTW~\cite{tobias2013youtube}              & 0.325 & 0.549 & 0.184 & 0.082 & 0.285 & 0.197 \\
      Xue2014~\cite{xue2014assessing}           & ---   & 0.778 & 0.388 & 0.219 & 0.462 & 0.360 \\
      Liu2012~\cite{liu2012case}                & 0.438 & 0.732 & 0.598 & 0.468 & 0.559 & 0.539 \\
      Yin2015~\cite{yin2015control}             & 0.441 & 0.686 & 0.741 & 0.541 & 0.602 & 0.601 \\
      VideoATLAS~\cite{bampis2017atlas}         & 0.076 & 0.673 & 0.469 & 0.670 & 0.472 & 0.603 \\
      Spiteri2016~\cite{spiteri2016bola}        & 0.493 & 0.711 & \textbf{0.798} & 0.662 & 0.662 & 0.680 \\
      P.1203~\cite{itu2017pnats}                & 0.415 & 0.821 & \textbf{0.797} & 0.668 & 0.675 & 0.708 \\
      Bentaleb2016~\cite{bentaleb2016sdndash}   & \textbf{0.650} & 0.883 & 0.718 & \textbf{0.692} & 0.735 & 0.730 \\
      SQI~\cite{duanmu2016sqi}                  & 0.644 & \textbf{0.906} & 0.690 & 0.690 & \textbf{0.735} & \textbf{0.732} \\ \hline
      KSQI                                      & \textbf{0.655} & \textbf{0.893} & 0.776 & \textbf{0.699} & \textbf{0.756} & \textbf{0.747} \\
    \bottomrule
    \end{tabular}
\end{table*}

\begin{table*}[t]
\centering
    \caption{KRCC between the objective QoE model prediction and MOS on the benchmark datasets.}\label{tab:krcc}
    \begin{tabular}{c|c c c c|c c}
    \toprule
      QoE model & LIVE-NFLX-I & LIVE-NFLX-II & WaterlooSQoE-III & WaterlooSQoE-IV & Average & Weighted Average\\ \hline
      Mok2011~\cite{mok2012qdash}               & 0.275 & 0.425 & 0.112 & 0.044 & 0.214 & 0.137 \\
      FTW~\cite{tobias2013youtube}              & 0.251 & 0.425 & 0.135 & 0.072 & 0.221 & 0.156 \\
      Xue2014~\cite{xue2014assessing}           & ---   & 0.582 & 0.262 & 0.148 & 0.148 & 0.253 \\
      Liu2012~\cite{liu2012case}                & 0.324 & 0.524 & 0.434 & 0.319 & 0.319 & 0.378 \\
      Yin2015~\cite{yin2015control}             & 0.327 & 0.482 & 0.543 & 0.379 & 0.379 & 0.427 \\
      VideoATLAS~\cite{bampis2017atlas}         & 0.050 & 0.491 & 0.330 & 0.480 & 0.338 & 0.432 \\
      Spiteri2016~\cite{spiteri2016bola}        & 0.376 & 0.501 & 0.597 & 0.461 & 0.484 & 0.490 \\
      P.1203~\cite{itu2017pnats}                & 0.300 & 0.619 & \textbf{0.604} & 0.479 & 0.501 & 0.520 \\
      Bentaleb2016~\cite{bentaleb2016sdndash}   & \textbf{0.479} & 0.712 & 0.521 & 0.495 & 0.552 & 0.538 \\
      SQI~\cite{duanmu2016sqi}                  & 0.475 & \textbf{0.735} & 0.496 & \textbf{0.504} & \textbf{0.553} & \textbf{0.543} \\ \hline
      KSQI                                      & \textbf{0.488} & \textbf{0.722} & \textbf{0.584} & \textbf{0.575} & \textbf{0.572} & \textbf{0.558} \\
    \bottomrule
    \end{tabular}
\end{table*}

\subsection{Experimental Setup}
\subsubsection{Objective QoE Models} We evaluate the performance of $11$ objective QoE models for adaptive streaming videos. The competing algorithms are chosen to cover a diversity of design philosophies, including $8$ classic parametric QoE models: FTW~\cite{tobias2013youtube}, Mok2011~\cite{mok2012qdash}, Liu2012~\cite{liu2012case}, Xue2014~\cite{xue2014assessing}, Yin2015~\cite{yin2015control}, Spiteri2016~\cite{spiteri2016bola}, Bentaleb2016~\cite{bentaleb2016sdndash}, and SQI~\cite{duanmu2016sqi}, $2$ state-of-the-art learning-based QoE models: VideoATLAS~\cite{bampis2017atlas} and P.1203~\cite{itu2017pnats}, and the proposed KSQI. A description of the existing QoE models is shown in Table~\ref{tab:existingQoE}. The implementation for VideoATLAS are obtained from the original authors and we implement the other nine QoE models. We have made the implementation of the models publicly available at \url{https://github.com/zduanmu/ksqi}. For the purpose of fairness, the parameters of all models are optimized on the WaterlooSQoE-I~\cite{duanmu2016sqi} and the WaterlooSQoE-II~\cite{duanmu2017qoe} datasets, except for P.1203~\cite{itu2017pnats} whose training methodology is not specified in the original paper. The WaterlooSQoE-I dataset contains $60$ compressed videos, $60$ compressed videos with initial buffering, and $60$ compressed videos with rebuffering. The WaterlooSQoE-II dataset involves $588$ video clips with variations in compression level, spatial resolution, and frame-rate. For the models with hyper-parameters, we randomly split the datasets into $80$\% training and $20$\% validation sets, and the hyper-parameters with the lowest validation loss are chosen. For KSQI, we set the maximum rebuffering duration $\tau_{\max}$ to $10$, because a rebuffering event longer than $10$ seconds rarely occurs as shown in Fig.~\ref{fig:feature_pdf}. The penalty of a rebuffering event longer than $10$ can be easily obtained by extrapolating the matrix $\mathbf{S}$. We set the step size $N$ to $10$, roughly characterizing the standard deviation of subjective presentation quality evaluation. The maximum presentation quality value $P=100$ is inherited from SSIMplus and VMAF. Although we can learn a initial buffering experience matrix independent from $\mathbf{S}$, it introduces unnecessary model complexity. Instead, we discount the impact of initial buffering with $\frac{1}{9}$ and set the expectation to the initial quality $P_{-1}$ to $80$ following the recommendation by~\cite{duanmu2016sqi}. We apply OSQP~\cite{stellato2017osqp} to solve the quadratic programming problem in~\eqref{eq:qp_s} and~\eqref{eq:qp_a}. The fidelity-smoothness tradeoff parameter $\lambda=1$ is obtained by cross-validation. In the subsequent section, we will also show that KSQI performs consistently over a broad range of $\lambda$ and step size $N$.

\subsubsection{Benchmark Databases}
We compare KSQI with state-of-the-art objective QoE models on four subject-rated adaptive streaming video datasets, including LIVE-NFLX-I~\cite{bampis2017qoe}, LIVE-NFLX-II~\cite{bampis2018towards}, WaterlooSQoE-III~\cite{duanmu2018quality}, and WaterlooSQoE-IV~\cite{duanmu2019sqoe4}. The LIVE-NFLX-I dataset consists of $112$ streaming videos derived from $14$ source content with $8$ handcrafted playout patterns. The LIVE-NFLX-II dataset consists of $420$ streaming videos generated from content-adaptive encoding profiles, bitrate adaptation algorithms and network conditions. The WaterlooSQoE-III dataset contains $450$ streaming videos of $20$ source contents recorded from a set of streaming experiment. The WaterlooSQoE-IV dataset contains $1,350$ highly-realistic streaming videos constructed from $5$ video contents, $2$ video encoders, $9$ real-world network traces, $5$ ABR algorithms, and $3$ viewing devices. The streaming videos in different datasets are of diverse characteristics since they are generated from different source videos, encoding profiles, adaptive streaming algorithms, and network conditions. We do not evaluate Xue2014 on the LIVE-NFLX-I dataset because their quantization parameters (QP) and encoded representations of the streaming videos are not publicly available.

\subsubsection{Evaluation Criteria} Three criteria are employed for performance evaluation by comparing MOS and objective QoE scores according to the recommendation by the video quality experts group~\cite{vqeg2000metrics}. We adopt Pearson linear correlation coefficient (PLCC) to evaluate the prediction accuracy, Spearman ranking-order correlation coefficient (SRCC) and Kendell rank correlation coefficient (KRCC) to assess prediction monotonicity. A better objective QoE model should have higher PLCC, SRCC, and KRCC.

\begin{figure}[t]
\centering
\includegraphics[width=0.9\linewidth]{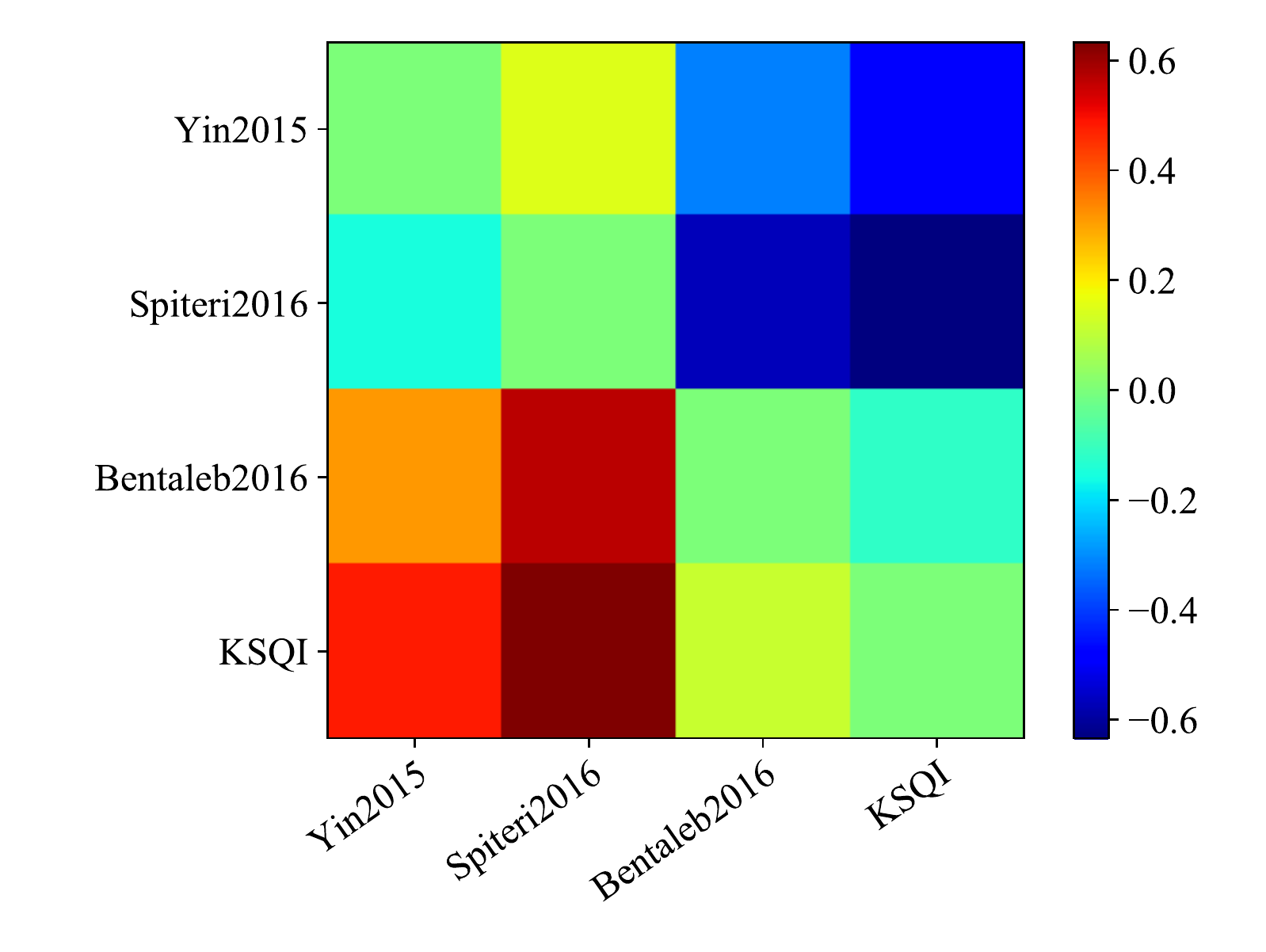}
\caption{Pairwise comparison matrix $\mathbf{R}$. Each entry indicates the preference of the row model against the column model. $\mathbf{R}-\mathbf{R}^T$ are drawn here for better visibility.}
\label{fig:pc_result}
\end{figure}

\subsection{Performance Comparison}
Tables~\ref{tab:plcc},~\ref{tab:srcc}, and~\ref{tab:krcc} show the PLCC, SRCC, and KRCC on the benchmark datasets, respectively, where top 2 best performers are highlighted with bold face. We have several observations. First, the objective QoE models which employ advanced VQA models as the presentation quality measure generally performs favorably against the conventional bitrate-based QoE models. In particular, Bentaleb2016 significantly outperforms Yin2015, where the only difference between them is the presentation quality measure. Second, although the learning-based QoE models perform competitively on certain test sets, they fail miserably on the other benchmark datasets. Specifically, the performance degradation of P.1203 and VideoATLAS from one dataset to another can be as large as 0.406 and 0.575, suggesting that the learning-based models exhibit low generalizability to diverse streaming environments. By contrast, KSQI achieves state-of-the-art performance on all benchmark datasets, thanks to the constraints given by domain knowledge. Third, the classic QoE models with a fixed parametric form cannot faithfully capture the subjective QoE response on streaming videos with complex distortion patterns, evident by the low prediction accuracy on WaterlooSQoE-III. In spite of the authors' effort in designing functional forms to conform known HVS properties~\cite{tobias2013youtube,xue2014assessing,duanmu2016sqi}, the QoE functions can vary significantly from exponential and logarithmic functions. On the other hand, KSQI does not assume a particular form of QoE functions and instead maximizes the mathematically well-behaveness. In summary, we believe the performance improvement arises because 1) KSQI is equipped with an HVS inspired VQA measure that generalizes well on a variety of video contents, encoders, and viewing devices; 2) the training procedure optimizes the quality prediction accuracy regularized by the prior knowledge on HVS; and 3) the proposed model does not make inaccurate a priori assumptions on the form of QoE functions.

\subsection{Best-case Validation}
Objective QoE model is not only used to evaluate, but also to optimize a variety of ABR algorithms and systems. A good rule of thumb is that an optimized system is only as good as the optimization criterion used to design it~\cite{wang2009mean}. Conversely, the performance of an objective QoE model can be assessed via synthesizing optimal streaming videos with respect to an objective QoE model followed by visual inspection of the generated stimulus~\cite{wang2004image,wang2008maximum}. Specifically, given a set of encoded and segmented videos and a realistic network trace, we can generate an optimal streaming video in terms of each objective QoE model. Subjective evaluation of the synthesized stimuli provides a best-case validation of the underlining objective QoE models. A good objective QoE model should produce perceptually better streaming videos comparing to the other schemes.

\begin{figure}[t]
\centering
\includegraphics[width=0.9\linewidth]{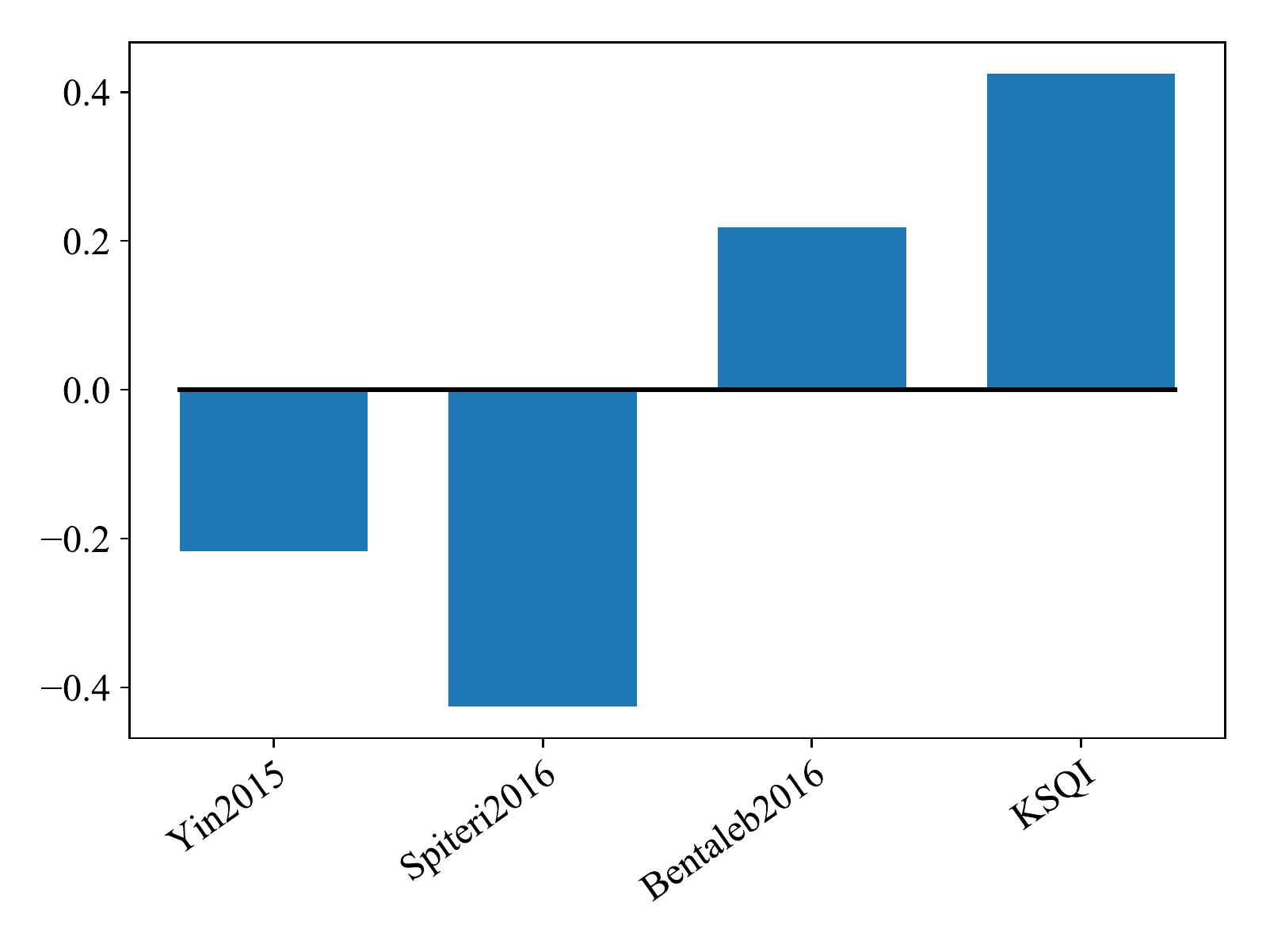}
\caption{Global ranking results of the four QoE models.}
\label{fig:pc_ag}
\end{figure}

\begin{table*}[t]
\centering
    \caption{Statistical significance matrix based on F-statistics on the combination of WaterlooSQoE-III, WaterlooSQoE-IV, LIVE-NFLX-I, and LIVE-NFLX-II datasets. A symbol ``1'' means that the performance of the row model is statistically better than that of the column model, a symbol ``0'' means that the row model is statistically worse, and a symbol ``-'' means that the row and column models are statistically indistinguishable.}\label{tab:vartest}
    \begin{tabular}{c|c c c c c c c c c c}
    \toprule
      & FTW & Mok2011 & Liu2012 & Yin2015 & VideoATLAS & Spiteri2016 & P.1203 & Bentaleb2016 & SQI & KSQI \\ \hline
      FTW               & - & - & 0 & 0 & 0 & 0 & 0 & 0 & 0 & 0  \\
      Mok2011           & - & - & 0 & 0 & 0 & 0 & 0 & 0 & 0 & 0  \\
      Liu2012           & 1 & 1 & - & - & 0 & 0 & 0 & 0 & 0 & 0  \\
      Yin2015           & 1 & 1 & - & - & 0 & 0 & 0 & 0 & 0 & 0  \\
      VideoATLAS        & 1 & 1 & 1 & 1 & - & 0 & 0 & 0 & 0 & 0  \\
      Spiteri2016       & 1 & 1 & 1 & 1 & 1 & - & - & 0 & 0 & 0  \\
      P.1203            & 1 & 1 & 1 & 1 & 1 & - & - & 0 & 0 & 0  \\
      Bentaleb2016      & 1 & 1 & 1 & 1 & 1 & 1 & 1 & - & 0 & 0  \\
      SQI               & 1 & 1 & 1 & 1 & 1 & 1 & 1 & 1 & - & 0  \\
      KSQI              & 1 & 1 & 1 & 1 & 1 & 1 & 1 & 1 & 1 & -  \\
    \bottomrule
    \end{tabular}
\end{table*}

In this paper, we select $12$ high-quality videos of diverse complexity to constitute the test sample set. All videos have the length of $30$ seconds. Using the source sequences, each video is encoded with an x264 encoder into $10$ representations in accordance with the Netflix's recommendation~\cite{netflix2015bitrateladder}. We segment the test sequences the encoded videos with GPAC's MP4Box~\cite{le2007gpac} with a segment length of $2$ seconds for the following reasons. First, $2$-second segments are widely used in the development of ABR algorithms. Second, it allows us to derive test videos in an efficient way such that they cover a diverse adaptation patterns in a limited time. $12$ network traces of diverse characteristics are randomly selected from the HSDPA dataset~\cite{riiser2013hdspa}. We compare KSQI with three objective QoE models that have guided the development of ABR algorithms, including Yin2015, Spiteri2016, and Bentaleb2016. We present results for the offline optimal scheme~\cite{spiteri2016bola,mao2017neural}, which is computed using dynamic programming with complete future throughput information. The dynamic programming-based method generates globally optimal streaming videos for the considered QoE models, completely eliminating the influence of inaccurate throughput estimation. For each source video, we randomly select a network trace and optimize the streaming videos with respect to the four objective QoE models. In the end, we obtain a total of $48$ streaming videos generated from $12$ (source videos, network traces) pairs $\times$ $4$ ABR algorithms. An online demonstration of the experiment is available at~\cite{duanmu2019sqoepc}.

We perform a subjective user study that adopts the pairwise comparison methodology in which a pair of streaming videos generated from the same video contents and network traces are presented to human viewers. The subjective experiment is setup as a normal indoor home settings with an ordinary illumination level, with no reflecting ceiling walls and floors. A customized interface is created to render a pair of $1920\times1080$ videos side-by-side on a 27 inch 4K monitor. The display is calibrated in accordance with the recommendations of ITU-R BT. 500~\cite{bt500subjective}. For each video pair, the subjects are forced to choose which one has a better perceptual quality. A total of $10$ na\"ive subjects, including $4$ males and $6$ females aged between $18$ and $35$, participate in the subjective experiment. Visual acuity and color vision are confirmed from each subject before the subjective test. A training session is performed, during which, 3 video pairs that are different from the videos in the testing set are presented to the subjects. We used the same methods to generate the videos used in the training and testing sessions. Therefore, subjects knew what distortion types would be expected before the test session, and thus learning effects are kept minimal in the subjective experiment. For each subject, the whole study takes one hour, which is divided into two sessions with a $5$-minute break in-between.

The results of the subjective experiment can be summarized as a $4$ $\times$ $4$ matrix $\mathbf{R}$, where $r_{i, j}$ represents the probability of QoE model $i$ better than QoE model $j$. Fig.~\ref{fig:pc_result} shows the result matrix $\mathbf{R}$, where the higher value of an entry (warmer color), the stronger the row model against the column model. It is obvious that KSQI performs favorably to the competing models. We further aggregate the pairwise comparison results into a global ranking via the maximum likelihood method for multiple options~\cite{tsukida2011analyze,ponomarenko2015image,ma2019gmad}. Let $\boldsymbol{\mu} = [\mu_1, \mu_2, \mu_3, \mu_4] \in \mathbb{R}^4$ be the global ranking score vector, we maximize the log-likelihood of $\boldsymbol{\mu}$
\begin{equation*}
    \begin{array}{ll}
    \underset{\boldsymbol{\mu}}{\argmax}
& \sum\limits_{i,j} r_{i, j}\log(\Phi(\mu_i - \mu_j)) \\
\text{subject to}
& \sum\limits_i \mu_i = 0,
    \end{array}
\end{equation*}
where $\Phi(\cdot)$ is the standard normal cumulative distribution function. The constraint $\sum_i \mu_i = 0$ is introduced to resolve the translation ambiguity. The optimization problem is convex and enjoys efficient solvers. A larger $\mu_i$ means the optimal streaming video in terms of the $i$-th model is perceptually better than the optimal samples generated by other QoE models in general. Fig.~\ref{fig:pc_ag} shows the experimental results. It can be seen that KSQI significantly outperforms the standard QoE models. The results have significant implications on the development of ABR algorithms. Specifically, state-of-the-art ABR algorithms have achieved a performance plateau levels and significant improvement has become difficult to attain. However, the enormous difference in perceptual relevance between the bitrate-based QoE model and KSQI suggests that further improvement is attainable simply by adopting perceptually motivated optimization criterion.

\begin{table*}[t]
\centering
    \caption{PLCC between the variants of KSQI prediction and MOS on the benchmark datasets.}\label{tab:abl1}
    \begin{tabular}{c|c c c c|c c}
    \toprule
      QoE model & LIVE-NFLX-I & LIVE-NFLX-II & WaterlooSQoE-III & WaterlooSQoE-IV & Average & Weighted Average \\ \hline
      KSQI with bitrate         & 0.622 & 0.722 & 0.670 & 0.618 & 0.658 & 0.647 \\
      KSQI with log bitrate     & 0.686 & 0.715 & 0.787 & \textbf{0.738} & 0.732 & 0.741 \\
      KSQI with QP              & ---   & 0.776 & 0.416 & 0.184 & 0.459 & 0.343 \\ \hline
      KSQI with VMAF            & \textbf{0.753} & \textbf{0.905} & \textbf{0.794} & 0.720 & \textbf{0.793} & \textbf{0.769} \\
    \bottomrule
    \end{tabular}
\end{table*}

\begin{table*}[t]
\centering
    \caption{PLCC between the variants of KSQI prediction and MOS on the benchmark datasets.}\label{tab:abl2}
    \begin{tabular}{c|c c c c|c c}
    \toprule
      Constraint \# & LIVE-NFLX-I & LIVE-NFLX-II & WaterlooSQoE-III & WaterlooSQoE-IV & Average & Weighted Average \\ \hline
      None                          & 0.731 & 0.903 & 0.663 & 0.681 & 0.745 & 0.720 \\
      \eqref{eq:s1}                 & 0.743 & 0.902 & 0.788 & 0.718 & 0.788 & 0.766 \\
      \eqref{eq:s1}\eqref{eq:s2}    & 0.748 & 0.904 & 0.780 & 0.719 & 0.788 & 0.765 \\
      \eqref{eq:s1}\eqref{eq:s2}\eqref{eq:s3} & 0.748 & 0.896 & \textbf{0.800} & 0.713 & 0.788 & 0.764 \\
      \eqref{eq:s1}\eqref{eq:s2}\eqref{eq:s3}\eqref{eq:s4} & \textbf{0.753} & \textbf{0.905} & 0.794 & \textbf{0.720} & \textbf{0.793} & \textbf{0.769} \\
      \eqref{eq:s1}\eqref{eq:s2}\eqref{eq:s3}\eqref{eq:s4}\eqref{eq:a1} & \textbf{0.753} & \textbf{0.905} & 0.794 & \textbf{0.720} & \textbf{0.793} & \textbf{0.769} \\
      \eqref{eq:s1}\eqref{eq:s2}\eqref{eq:s3}\eqref{eq:s4}\eqref{eq:a1}\eqref{eq:a2}    & \textbf{0.753} & \textbf{0.905} & 0.793 & \textbf{0.720} & \textbf{0.793} & \textbf{0.769} \\ 
      \eqref{eq:s1}\eqref{eq:s2}\eqref{eq:s3}\eqref{eq:s4}\eqref{eq:a1}\eqref{eq:a2}\eqref{eq:a3}    & \textbf{0.753} & \textbf{0.905} & 0.794 & \textbf{0.720} & \textbf{0.793} & \textbf{0.769} \\
      \eqref{eq:s1}                 & 0.744 & 0.902 & 0.788 & 0.718 & 0.788 & 0.766 \\
      \eqref{eq:s2}                 & 0.743 & 0.906 & 0.758 & 0.717 & 0.781 & 0.760 \\
      \eqref{eq:s3}                 & 0.743 & 0.895 & 0.798 & 0.713 & 0.788 & 0.764 \\
      \eqref{eq:s4}                 & 0.753 & 0.902 & 0.787 & 0.717 & 0.790 & 0.766 \\
      \eqref{eq:a1}                 & 0.745 & 0.884 & 0.770 & 0.691 & 0.773 & 0.744 \\
      \eqref{eq:a2}                 & 0.745 & 0.884 & 0.770 & 0.692 & 0.773 & 0.744 \\
      \eqref{eq:a3}                 & 0.745 & 0.884 & 0.770 & 0.691 & 0.773 & 0.744 \\
      \eqref{eq:a4}                 & 0.746 & 0.884 & 0.770 & 0.692 & 0.773 & 0.744 \\ \hline
      KSQI                          & \textbf{0.753} & \textbf{0.905} & 0.794 & \textbf{0.720} & \textbf{0.793} & \textbf{0.769} \\
    \bottomrule
    \end{tabular}
\end{table*}

\subsection{Statistical Significance Test}
To ascertain that the improvement of the proposed model is statistically significant, we carry out a statistical significance analysis by following the approach introduced in~\cite{sheikh2006statistical}. First, we linearly scale MOSs in each dataset to the same perceptual scale [0, 100]. Second, a nonlinear regression function is applied to map the objective quality scores to predict the subjective scores independently on the four testing datasets. The prediction residuals of each QoE models from all datasets are aggregated into a vector. We observe that the prediction residuals all have zero-mean, and thus the model with lower variance is generally considered better than the one with higher variance. We conduct a hypothesis testing using F-statistics. Since the number of samples exceeds 50, the Gaussian assumption of the residuals approximately hold based on the central limit theorem~\cite{bishop2006pattern}. The test statistic is the ratio of variances. The null hypothesis is that the prediction residuals from one quality model come from the same distribution and are statistically indistinguishable (with 95\% confidence) from the residuals from another model. After comparing every possible pairs of objective models, the results are summarized in Table~\ref{tab:vartest}, where a symbol `1' means the row model performs significantly better than the column model, a symbol `0' means the opposite, and a symbol `-' indicates that the row and column models are statistically indistinguishable. It can be observed that the proposed model is statistically better than all other methods on the combination of all existing benchmark datasets.

\begin{figure}[t]
\centering
\includegraphics[width=0.9\linewidth]{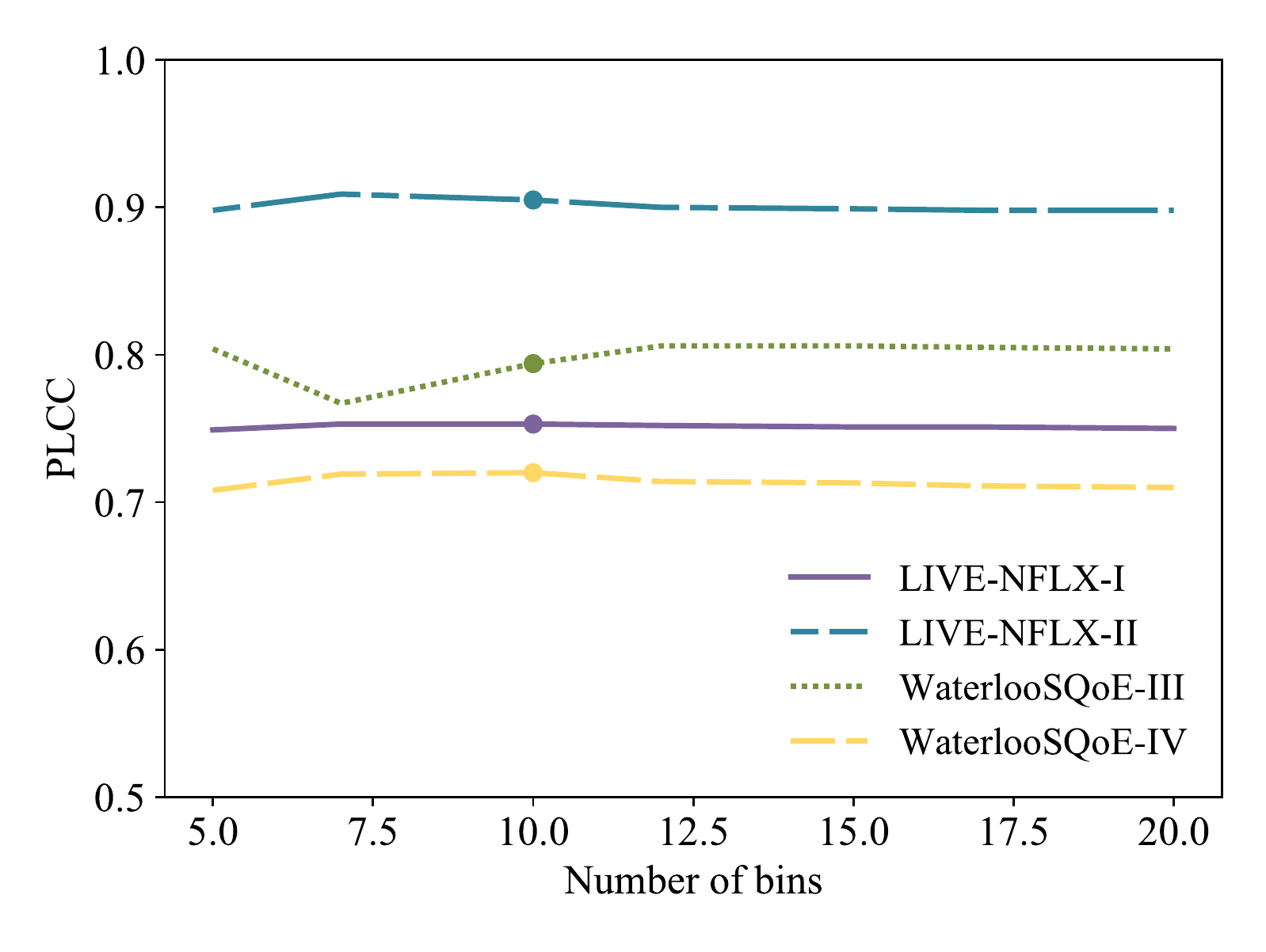}
\caption{Performance of KSQI with different number of bins.}
\label{fig:bin_num}
\end{figure}

\subsection{Ablation Experiment}
We conduct a series of ablation experiments to single out the core contributors of KSQI. We first take bitrate~\cite{liu2012case,yin2015control}, logarithmic bitrate~\cite{spiteri2016bola}, and QP~\cite{xue2014assessing} as the video presentation quality measure as opposed to VMAF and then train the QoE model with the proposed optimization framework. In order to map the range of video presentation quality measure into the same perceptual scale [0, 100], we apply a linear transform to the alternative measures before the training stage. From Table~\ref{tab:abl1}, we observe that KSQI achieves the best performance with the adoption of state-of-the-art video quality measure such as VMAF.

\begin{figure}[t]
\centering
\includegraphics[width=0.9\linewidth]{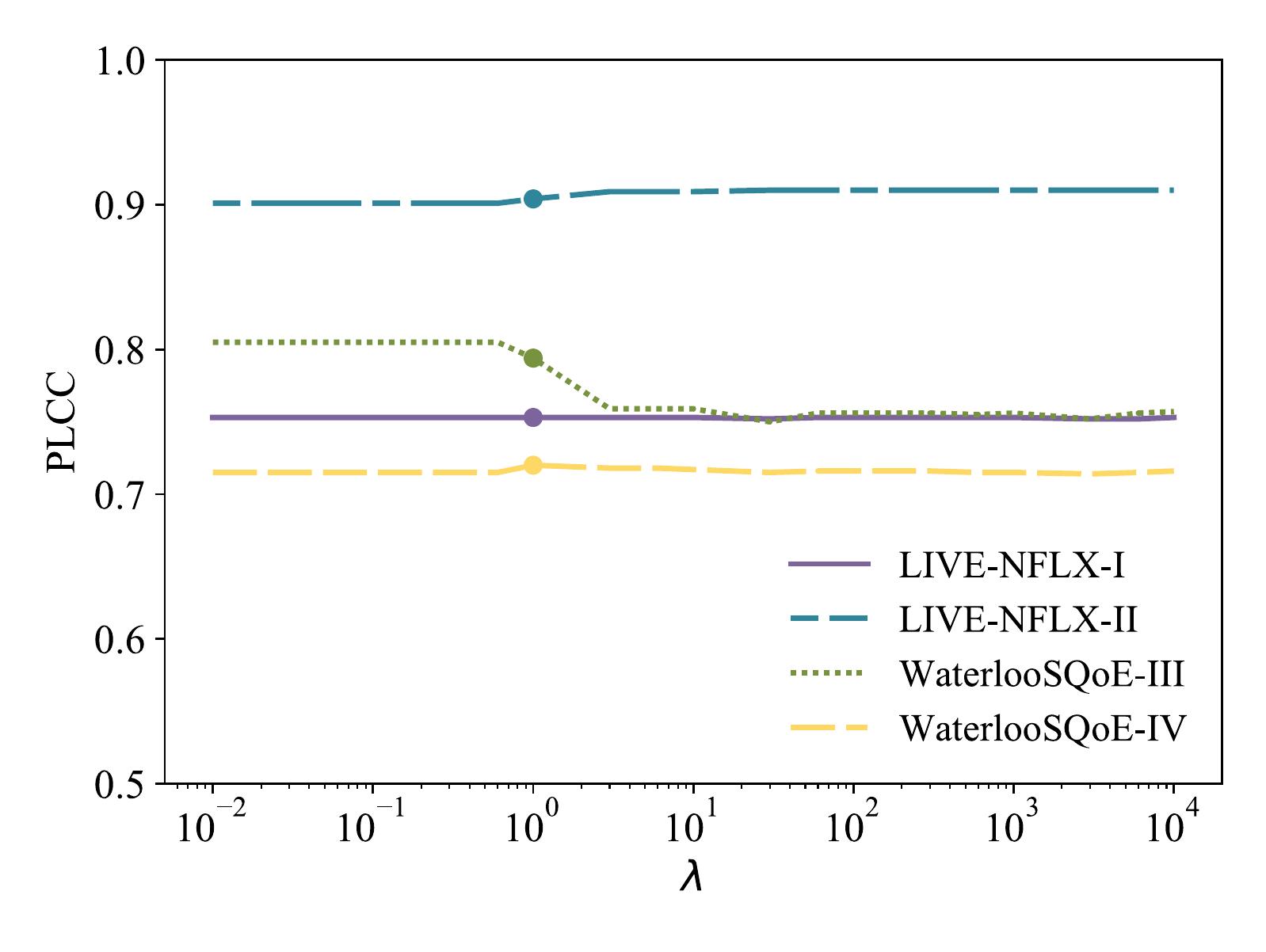}
\caption{Performance of KSQI with different $\lambda$.}
\label{fig:lambda}
\end{figure}

Next, we analyze the impact of the knowledge-imposed constraints on the quality prediction performance. We start from a baseline model by solving the problem in~\eqref{eq:qp_s} and~\eqref{eq:qp_a} with no constraints and gradually increase the number of constraints. We then investigate the validity of each observation by imposing only one constraint in a variant model. The results are listed in Table~\ref{tab:abl2}, from which the key observations are as follows. First, the performance of KSQI generally improves with respect to the number of imposed constraints, advocating the effectiveness of prior knowledge in regularizing the objective QoE functions. Second, while some of the constraints do not improve the performance of KSQI by themselves, the joint model achieves state-of-the-art performance. This suggests that the constraints may be complement to each other. Third, the constraint~\eqref{eq:s3} has drastically different impacts on the LIVE-NFLX-II dataset and the WaterlooSQoE-III dataset, suggesting that the validity of the constraint may be influenced by other factors. A careful investigation may further improve the performance of the proposed QoE model.

\subsection{Impact of Step Sizes}
In previous experiments, we set the bin sizes of video presentation quality and rebuffering duration to $10$ and $1$, respectively. To investigate the impact of step sizes, we train several variants of KSQI, where the number of bins ranges from $5$ to $20$. We show the experimental results in Fig.~\ref{fig:bin_num}. Theoretically speaking, the performance of KSQI should increase monotonically with respect to the precision of feature representations. However, the observation does not echo our expectation, which may be a consequence of insufficient training data and intrinsic noise in the subjective opinion scores. Nevertheless, KSQI is generally very robust to a broad range of bin sizes.

\subsection{Impact of $\lambda$}
The parameter $\lambda$ in KSQI determines the tradeoff between fidelity and smoothness of the QoE functions. Although the optimal parameter is obtained from cross-validation in previous experiments, we also perform an experiment to investigate the impact of $\lambda$. Specifically, we train several versions of KSQI, where $\lambda$ ranges from $0.01$ to $10,000$. The results are shown in Fig.~\ref{fig:lambda}, from which we can observe that the performance of KSQI is generally insensitive to $\lambda$.

\section{Conclusions}\label{sec:conclusion}
We propose a novel objective QoE model for adaptive streaming videos, namely KSQI, by regularizing a non-parametric model with known HVS properties. KSQI outperforms the existing objective QoE models by a sizable margin over a wide range of video contents, encoding configurations, network conditions, and viewing devices, which we believe arises from a perceptually motivated video quality representation, a knowledge constrained optimization framework, and a non-parametric model of QoE functions.

The proposed model may be improved in many ways. First, KSQI is readily extendable when new knowledge of HVS properties is acquired. With proper modifications of the non-parametric functions, we may incorporate more features such as motion strength~\cite{liu2015deriving} into the QoE model. Second, there may be better ways to combine the video presentation quality, rebuffering experience, and quality adaptation experience. For example, we can jointly model all influencing factors by escalating the dimensionality of the non-parametric model. Third, how to integrate the QoE model into the adaptive bitrate selection algorithm for optimal playback control is another challenging problem that is worth further investigations.

\bibliographystyle{IEEEtran}
\bibliography{ref} 

\end{document}